\documentstyle[11pt,a4]{article}

\begingroup\makeatletter\ifx\SetFigFont\undefined
\def\x#1#2#3#4#5#6#7\relax{\def\x{#1#2#3#4#5#6}}%
\expandafter\x\fmtname xxxxxx\relax \def\y{splain}%
\ifx\x\y   
\gdef\SetFigFont#1#2#3{%
  \ifnum #1<17\tiny\else \ifnum #1<20\small\else
  \ifnum #1<24\normalsize\else \ifnum #1<29\large\else
  \ifnum #1<34\Large\else \ifnum #1<41\LARGE\else
     \huge\fi\fi\fi\fi\fi\fi
  \csname #3\endcsname}%
\else
\gdef\SetFigFont#1#2#3{\begingroup
  \count@#1\relax \ifnum 25<\count@\count@25\fi
  \def\x{\endgroup\@setsize\SetFigFont{#2pt}}%
  \expandafter\x
    \csname \romannumeral\the\count@ pt\expandafter\endcsname
    \csname @\romannumeral\the\count@ pt\endcsname
  \csname #3\endcsname}%
\fi
\fi\endgroup
 
\makeatletter

\@addtoreset{equation}{section}
\def\section{\@startsection {section}{1}{\z@}{-3.5ex plus -1ex minus
 -.2ex}{2.3ex plus .2ex}{\large\bf\centering}}
\def\subsection{\@startsection{subsection}{2}{\z@}{-3.25ex plus%
 -1ex minus -.2ex}{1.5ex plus .2ex}{\bf}}
\def\subsubsection{\@startsection{subsubsection}{3}{\z@}{-3.25ex plus%
 -1ex minus -.2ex}{1.5ex plus .2ex}{\sl}}
\makeatother
 
\newcommand{\st}{{\rm spin}}
\newcommand{\nn}{\nonumber\\}
\newcommand{\uqgt}{\widetilde{{\cal U}_q(\hat{g})}}
\newcommand{\uqg}{{\cal U}_q({\hat g})}
\newcommand{\psa}{\pi}
\newcommand{\D}{\Delta}
\newcommand{\sabd}{S_{ab}(\theta -\theta')}
\newcommand{\pat}{\pi_{a}^{(\theta)}}
\newcommand{\psat}{\pi_{s,a}^{(\theta)}}
\newcommand{\psbt}{\pi_{s,b}^{(\theta)}}
\newcommand{\qgta}{{\cal U}_q(g_2^{(1)})}
\newcommand{\qgt}{{\cal U}_q(g_2)}
\newcommand{\qat}{{\cal U}_q(a_2)}
\newcommand{\qgtat}{\widetilde{{\cal U}_q(\hat{g_2})}}
\newcommand{\qt}{\sqrt{\frac{q^2+1}{q}}}
\newcommand{\R}{{\widehat{\cal R}}(x,q)}
\def\ampl#1{\frac{\displaystyle x-q^{#1}}{\displaystyle 1-xq^{#1}}}
\def\rampl#1{\frac{\displaystyle 1-xq^{#1}}{\displaystyle x-q^{#1}}}
\def\rf#1{(\ref{#1})}
\newcommand{\repvect}[3]{ | {\bf #1}\ ;\ #2 ,  #3\rangle}
\newcommand{\id}{{\rm I}}
\newcommand{\ggblock}[1]{\left(#1\right)}

\begin{document}
\vbox{\begin{flushright}
ITP Budapest Report No. 525\\
Revised on 1/3/1997
\end{flushright}
\title{Quantum Affine Symmetry and Scattering Amplitudes of the 
Imaginary Coupled $d_4^{(3)}$ Affine Toda Field Theory}  
\author{ G\'abor Tak\'acs \\
         Institute for Theoretical Physics\\
         E\"otv\"os University \\
         Budapest, Hungary}
\date{23rd January 1996}
\maketitle }
\begin{abstract}
An exact $S$-matrix is conjectured for the imaginary coupled $d_4^{(3)}$ 
affine Toda field theory, using the $\qgta$ symmetry. It is shown that 
this $S$-matrix is consistent with the results for the case of real 
coupling using the breather-particle correspondence. For $q$ a root of 
unity it is argued that the theory can be restricted to yield 
$\Phi(11|14)$ perturbations of $WA_2$ minimal models and the 
restriction is performed for the $(3,p')$ minimal models. 
\end{abstract}
\section{Introduction}

Affine Toda field theories form a very important class of
two-dimensional integrable field theories (for a review see
\cite{corrigan}). 
In the imaginary coupling case, these 
theories provide natural generalizations of sine-Gordon theory and 
have solitonic excitations in their spectra. While in general these
models are nonunitary as quantum field theories, their RSOS
restrictions  
correspond to perturbations of W-symmetric rational conformal field 
theories (RCFTs), among them to unitary ones \cite{Wrestr}. 

In the case of theories associated to simply-laced affine Lie algebras, 
the semi-classical mass ratios are stable under quantum 
corrections \cite{hollowood}
and the $S$-matrices can be obtained using the fact 
that the theories are invariant under a quantum affine symmetry algebra 
of nonlocal charges. In the nonsimply-laced case, while the mass 
ratios are not stable under quantum corrections \cite{unstable}, 
it is again thought that the $S$-matrix can be obtained using the 
representation theory of the nonlocal symmetry algebra. 
We remark that the existing computations 
of the mass renormalization \cite{unstable} in the nonsimply-laced case 
do not agree with each other and also that the instability of the 
classical solutions \cite{khastgir} casts a big question mark over 
the validity of the results in \cite{hollowood,unstable}. 
However, it is still plausible that the mass ratios of the
nonsimply-laced 
theories would be changed by quantum corrections similarly 
to the real coupling theories \cite{braden}, but it is unclear how to
make a 
consistent semiclassical quantisation in the imaginary coupling case.

A nice review of the concept 
of applying the quantum symmetry algebra to construct exact $S$-matrices 
can be found in \cite{delius}. 
In several cases the exact $S$-matrices have been 
computed: for $a_n^{(1)}$ affine Toda theory \cite{An1smat}, for 
the $d_n^{(2)}$ \cite{Dn2smat} and the $b_n^{(1)}$ \cite{Bn1smat} case.

In this paper we will treat 
the imaginary coupled $d_4^{(3)}$ affine Toda field theory. The 
restrictions of this theory are argued to be connected with 
certain perturbations of $WA_2$ minimal models.

The layout of the paper is as follows: in Section 2 we briefly 
review the known facts about the quantum symmetry and the $S$-matrix. 
The derivation of the S-matrix is described in Section 3, together 
with an analysis of the bound state poles and the connection 
with the real coupling case via the breather-particle correspondence. 
Section 4 is devoted to the restriction to perturbed $WA_2$ minimal 
models, while in Section 5 we draw our conclusions. The paper 
ends with two Appendices (A and B), containing some formulas 
used in the main text.

\section{Quantum symmetry and $S$-matrix}

\subsection{The quantum affine symmetry}

Let us take an affine Lie algebra $\hat g$ and define the affine 
Toda field theory with the Lagrangian
\begin{equation}
S=\int d^2x \frac{1}{2}\partial_\mu\vec{\Phi}\partial_\mu\vec{\Phi}
+\frac{\lambda}{2\pi}\int d^2x \sum\limits_{\vec{\alpha}_j}
\exp\left(i\beta\frac{2}{(\vec{\alpha}_j,\vec{\alpha}_j)}
\vec{\alpha}_j\cdot\vec{\Phi}\right)\ ,
\label{atftlagr}\end{equation}
where the vectors $\vec{\alpha}_j\ ,\ j=0\dots r$ are the simple roots 
of $\hat g$ (meaning the simple roots of $g$ plus the extending or
affine 
root with label $0$). The normalization of the roots is given by taking 
$(\vec{\alpha}_j,\vec{\alpha}_j)=2$ for the long roots.

In the usual nomenclature, (\ref{atftlagr}) is referred to as the 
${\hat g}^\vee$ affine Toda action, where ${\hat g}^\vee$ denotes 
the affine Lie algebra dual to $\hat g$, 
whose roots $\vec{\gamma}_j$ are the coroots of $\hat g$
\begin{equation}
\vec{\gamma}_j=\frac{2\vec{\alpha}_j}{(\vec{\alpha}_j,\vec{\alpha}_j)}
\ .
\end{equation} 
It is immediately apparent that any simply-laced affine Lie algebra is 
self-dual, while for nonsimply-laced ones the dual is obtained by 
reversing the arrows in the Dynkin diagram. Our convention of the action 
follows \cite{feldlecl} and is different from the one usually 
adopted in the literature. This was chosen for later convenience.

The theory (\ref{atftlagr}) is known to be integrable. Besides the 
infinite number of commuting charges, however, there is a nonlocal 
non-abelian symmetry algebra, which is given by the quantum symmetry 
algebra ${\cal U}_q({\hat g})$ \cite{feldlecl,berlecl}. 
The parameter $q$ is related to 
the coupling constant by 
\begin{equation}
q=\exp\left(\frac{4\pi^2 i}{\beta^2}\right)\ .
\end{equation}
For later convenience, we introduce another parametrization of the
coupling 
constant:
\begin{equation}
\xi = \frac{\pi\beta^2}{8\pi-3\beta^2}\ ,\ 
q=\exp\left[i\pi\left(\frac{\pi}{2\xi}+\frac{3}{2}\right)\right]\ .
\end{equation}
We suppose that $0<\xi<\infty$, which means $0<\beta^2<8\pi/3$. In 
analogy with sine-Gordon theory, the point $\beta^2=8\pi/3$ must 
be the coupling at which the interaction terms become irrelevant 
and the theory describes two free scalar fields at the 
infrared when $\beta^2>8\pi/3$. We will see later that this 
is consistent with the exact $S$-matrix and also with arguments 
using perturbed conformal field theory.

The restriction of the above theory to perturbed RCFT occurs when the 
parameter $q$ is a root of unity and leads to the so-called RSOS 
scattering amplitudes. 

To fix our conventions of the quantum affine Lie algebra, let me 
briefly summarize the defining relations. We define first 
the algebra $\uqgt$, which is generated 
by elements $\{ h_i,\ e_i,\ f_i,\ i=0\dots r\}$, satisfying the 
following commutation relations
\begin{eqnarray}
&&\left[h_i,h_j\right]=0,\ \left[h_i,e_j\right]=(\alpha_i,\alpha_j)e_j,\ 
\left[h_i,f_j\right]=-(\alpha_i,\alpha_j)f_j,\nonumber\\
&&\left[e_i,f_j\right]=\delta_{ij}\frac{q^{h_i}-q^{-h_i}}{q_i-q_i^{-1}},\ 
q_i=q^{(\alpha_i,\alpha_i)/2}\ ,
\end{eqnarray}
together with the quantum Serre relations
\begin{eqnarray}
\sum\limits_{k=0}^{1-a_{ij}}(-1)^k\left(\matrix{1-a_{ij}\cr
k}\right)_{q_i}
e_i^ke_je_i^{1-a_{ij}-k}=0\ ,\nn
\sum\limits_{k=0}^{1-a_{ij}}(-1)^k\left(\matrix{1-a_{ij}\cr
k}\right)_{q_i}
f_i^kf_jf_i^{1-a_{ij}-k}=0\ ,\nn 
i\neq j\ ,
\end{eqnarray}
where
\begin{equation}
\left(\matrix{m\cr k}\right)_{q_i}=\frac{[m]_q!}{[k]_q![m-k]_q!},\ 
[m]_q!=\prod\limits_{1\leq i\leq m}[i]_q,\
[i]_q=\frac{q^i-q^{-i}}{q-q^{-1}}
\end{equation}
are the usual quantum binomial coefficients and
\begin{equation}
a_{ij}=\frac{2(\alpha_i,\alpha_j)}{(\alpha_j,\alpha_j)}
\end{equation}
is the Cartan matrix of $g_2^{(1)}$. The coproduct is given by 
\begin{eqnarray}
&&\Delta (e_i)=q^{-h_i/2}\otimes e_i+e_i\otimes q^{h_i/2}\ ,\nn
&&\Delta (f_i)=q^{-h_i/2}\otimes f_i+f_i\otimes q^{h_i/2}\ ,\nn
&&\Delta(q^{\pm h_i/2})=q^{\pm h_i/2}\otimes q^{\pm h_i/2}\ .
\end{eqnarray}
The conserved charges possess a definite Lorentz spin.
Denoting the infinitesimal two-dimensional Lorentz generator by $D$ we
have
\begin{equation}
\label{derivation}
[D,e_i]= s_i e_i,~~[D,f_i]=- s_i f_i,
~~[D,H_i]=0,~~i=0,\dots,r.
\end{equation}
where $s_i$ is the Lorentz spin of $e_i$. Adjoining the operator $D$ to 
the algebra $\uqgt$ results in the full algebra $\uqg$. 

Denoting the Lorentz spin of an operator $A$ by $\st(A)$, 
$\st:~\uqg \rightarrow {\rm R}$ is a gradation of $\uqg$, which is 
uniquely 
fixed by giving $s_0,\dots,s_r$. The change between the gradations can 
be performed with similarity transformations by exponentials of the 
Cartan elements $h_i$. 

Denote the one-particle states by $|a,\alpha,\theta\rangle$, where $a$ 
denotes the multiplet, $\alpha$ is the label within the multiplet and 
$\theta$ is the rapidity.
The rapidity specifies the energy $E=m\, \mbox{cosh}(\theta)$ and the
momentum
$p=m\,\mbox{sinh}(\theta)$, where $m$ denotes the mass of the particle.
At fixed rapidity the particles in the multiplet $a$ span the space
$V_a$ which carries a finite dimensional unitary
representation $\pi_a$ of $\uqgt$, with zero central charge.
Including the rapidity the one-particle space will be denoted as 
$V_a(\theta)$.
Under a finite Lorentz transformation $L(\lambda)=\exp(\lambda D)$ the 
rapidity $\theta$ is shifted by $\lambda$
\begin{equation}\label{lorentz}
L(\lambda)|a,\theta\rangle=|a,\theta+\lambda\rangle.
\end{equation}
{}From this we deduce that $V_a(\theta )$ carries the following
infinite dimensional representation $\psa$ of $\uqg$:
\begin{eqnarray}
\pat(D)&=&\frac{d}{d\theta},\nn
\pat(e_i)&=&\pi_a(e_i)e^{s_i\theta},\nn
\pat(f_i)&=&\pi_a(f_i)e^{-s_i\theta},\nn
\pat(h_i)&=&\pi_a(h_i).
\end{eqnarray}

The action of the symmetry on asymptotic multi-particle states is given 
by the coproduct:
\begin{equation}\label{multiact}
\pi_{a_1\cdots a_n}^{(\theta_1\cdots\theta_n)}(A)
=(\pi_{a_1}^{(\theta_1)}\otimes\cdots\otimes\pi_{a_n}^{(\theta_n)})
\D^{n-1}(A),
\end{equation}
where $\D^2=(1\otimes\D)\D,~~\D^3=(1\otimes 1\otimes\D)\D^2$, etc.

\subsection{The two-particle S-matrices}

The $S$-matrix gives the mapping of an incoming asymptotic
two-particle state into an outgoing asymptotic two-particle state
\begin{equation}
 \sabd:~V_a(\theta)\otimes V_b(\theta')\rightarrow
V_b(\theta')\otimes V_a(\theta)
\end{equation}
\begin{equation}
|b,\beta', \theta'\rangle\otimes|a,\alpha', \theta\rangle=
\left( \sabd\right)_{\alpha\beta}^{\alpha'\beta'}
\left(|a,\alpha, \theta\rangle\otimes|b,\beta, \theta'\rangle\right)
\end{equation}
The quantum affine symmetry tells us that
\begin{equation}\label{intertwine}
\sabd\pi_{ab}^{(\theta\theta')}(A)=\pi_{ba}^{(\theta'\theta)}(A)\sabd,
{}~~~~~\forall A\in\uqgt.
\end{equation}
\rf{intertwine}  means that $\sabd$ is an intertwiner between the
representation $\pi_{ab}^{(\theta\theta')}$
and the representation $\pi_{ba}^{(\theta'\theta)}$. 
Because these representations
are irreducible for generic $\theta$, $\theta'$, such an intertwiner is 
unique, up to an overall constant. This intertwiner is obtained by
evaluating the universal R-matrix of $\uqgt$ in the appropriate
representation and gradation
\begin{equation}
\check{R}^{(s)}_{ab}(\theta-\theta')=
P_{ab}\left((\psat\otimes\psbt) R\right)
\end{equation}
and multiplying it by an overall scalar prefactor $f_{ab}$,
\begin{equation}\label{sfromr}
\sabd=f_{ab}(\theta-\theta') \check{R}^{(s)}_{ab}(\theta-\theta').
\end{equation}
Here $P_{ab}:~V_a(\theta)\otimes V_b(\theta')\rightarrow
V_b(\theta')\otimes V_a(\theta)$ is the permutation operator
$P_{ab}: V_a\otimes V_b\mapsto V_b\otimes V_a$.
The prefactor $f_{ab}(\theta)$ will be constrained by the requirements
of
unitarity, crossing symmetry and the bootstrap principle.

By definition, the universal R-matrix of $\uqgt$ satisfies
\begin{equation}\label{rdef}
R\D(A)=\D^{op}(A)R~~~~~\forall A\in\uqgt,
\end{equation}
where $\D^{op}$ is the opposite coproduct obtained by interchanging the
factors of the tensor product. 

It can further be derived that multi-particle S-matrices are given 
by products of two-particle ones, the consistency of which is provided 
by the Yang-Baxter equation fulfilled by the universal R-matrix. 

\section{The $S$-matrix of the fundamental solitons in imaginary coupled 
$d_4^{(3)}$ Toda theory}

The quantum symmetry of the $d_4^{(3)}$ theory is given by $\qgta$. 
The Cartan matrix is
\begin{equation}
\left(\matrix{ 2 & -1 &  0 \cr 
              -1 &  2 & -1 \cr
               0 & -3 &  2 }\right)
\end{equation}
and the simple roots are given by 
\begin{eqnarray}
&&\alpha_0=(-1/\sqrt{2},-\sqrt{3/2})\ , \nn 
&&\alpha_1=(\sqrt{2},0)\ ,\ 
\alpha_2=(-1/\sqrt{2},1/\sqrt{6})\ .
\end{eqnarray}
There are two important subalgebras of $\qgta$: the generators 
$\{h_i,e_i,f_i,i=1,2\}$ form a subalgebra ${\cal A}_1$ isomorphic to 
$\qgt$, while the algebra ${\cal A}_0$ generated by 
$\{h_i,e_i,f_i,i=0,1\}$ is isomorphic to $\qat$.

\subsection{The R-matrix in the fundamental representation}

We will assume that the fundamental solitons transform in the 
$7$-dimensional representation of the algebra, which is also the 
fundamental representation of ${\cal A}_1$. In this space, the algebra 
$\qgtat$ is represented by the following matrices:
\begin{eqnarray}
&&h_0=\left(\matrix{ 
             -1& 0 & 0 & 0 & 0 & 0 & 0 \cr
             0 &-1 & 0 & 0 & 0 & 0 & 0 \cr
             0 & 0 & 0 & 0 & 0 & 0 & 0 \cr
             0 & 0 & 0 & 0 & 0 & 0 & 0 \cr
             0 & 0 & 0 & 0 & 0 & 0 & 0 \cr
             0 & 0 & 0 & 0 & 0 & 1 & 0 \cr
             0 & 0 & 0 & 0 & 0 & 0 & 1 }\right),
h_1=\left(\matrix{ 
             0 & 0 & 0 & 0 & 0 & 0 & 0 \cr
             0 & 1 & 0 & 0 & 0 & 0 & 0 \cr
             0 & 0 &-1 & 0 & 0 & 0 & 0 \cr
             0 & 0 & 0 & 0 & 0 & 0 & 0 \cr
             0 & 0 & 0 & 0 & 1 & 0 & 0 \cr
             0 & 0 & 0 & 0 & 0 &-1 & 0 \cr
             0 & 0 & 0 & 0 & 0 & 0 & 0 }\right),\nn 
&&h_2=\frac{1}{3}\left(\matrix{ 
             1 & 0 & 0 & 0 & 0 & 0 & 0 \cr
             0 &-1 & 0 & 0 & 0 & 0 & 0 \cr
             0 & 0 & 2 & 0 & 0 & 0 & 0 \cr
             0 & 0 & 0 & 0 & 0 & 0 & 0 \cr
             0 & 0 & 0 & 0 &-2 & 0 & 0 \cr
             0 & 0 & 0 & 0 & 0 & 1 & 0 \cr
             0 & 0 & 0 & 0 & 0 & 0 &-1 }\right),\nn
&&e_0=\left(\matrix{ 
             0 & 0 & 0 & 0 & 0 & 0 & 0 \cr
             0 & 0 & 0 & 0 & 0 & 0 & 0 \cr
             0 & 0 & 0 & 0 & 0 & 0 & 0 \cr
             0 & 0 & 0 & 0 & 0 & 0 & 0 \cr
             0 & 0 & 0 & 0 & 0 & 0 & 0 \cr
             1 & 0 & 0 & 0 & 0 & 0 & 0 \cr
             0 & 1 & 0 & 0 & 0 & 0 & 0 }\right),
e_1=\left(\matrix{ 
             0 & 0 & 0 & 0 & 0 & 0 & 0 \cr
             0 & 0 & 1 & 0 & 0 & 0 & 0 \cr
             0 & 0 & 0 & 0 & 0 & 0 & 0 \cr
             0 & 0 & 0 & 0 & 0 & 0 & 0 \cr
             0 & 0 & 0 & 0 & 0 & 1 & 0 \cr
             0 & 0 & 0 & 0 & 0 & 0 & 0 \cr
             0 & 0 & 0 & 0 & 0 & 0 & 0 }\right),\nn
&&e_2=\left(\matrix{ 
             0 & 1 & 0 & 0 & 0 & 0 & 0 \cr
             0 & 0 & 0 & 0 & 0 & 0 & 0 \cr
             0 & 0 & 0 &\qt& 0 & 0 & 0 \cr
             0 & 0 & 0 & 0 &\qt& 0 & 0 \cr
             0 & 0 & 0 & 0 & 0 & 0 & 0 \cr
             0 & 0 & 0 & 0 & 0 & 0 & 1 \cr
             0 & 0 & 0 & 0 & 0 & 0 & 0 }\right),\nn
&&f_i=e_i^{tr}\ ,\ i=0,1,2\ .
\end{eqnarray}
($^{tr}$ denotes usual matrix transposition.)
In the following we will use the principal gradation, in which case 
all the rapidity dependence is shifted to the generators with index $0$,
i.e.
\begin{equation}
\pi^{(\theta)}(h_i)=h_i\ ,\ 
\pi^{(\theta)}(h_i)=x^{\delta_{i0}}e_i\ ,\ 
\pi^{(\theta)}(h_i)=x^{-\delta_{i0}}f_i\ ,
\end{equation}
where $x$ is essentially the exponential of the rapidity (the precise 
correspondence will be given later).

We will solve the intertvining equation for the operator 
\begin{equation}
\R =P_{12}{\cal R}(x,q)\ ,
\end{equation}
where ${\cal R}(x,q)$ denotes the 
universal R-matrix in the tensor product of two fundamental
representations
and $x$ denotes the ratio $x_1/x_2$ of the spectral parameters in the
first 
and second space, respectively.

The equations for the generators $X\in {\cal A}_1$ look like
\begin{equation}
[\R,\Delta( X)]=0\ .
\end{equation} 
This means that $\R$ is a $\qgt$ invariant operator in the space 
${\bf 7}\otimes {\bf 7}$. This space decomposes into irreducible 
representations under $\qgt$ in the following way
\begin{equation}
{\bf 7}\otimes {\bf 7}={\bf 1}\oplus {\bf 7}\oplus {\bf 14}\oplus {\bf
27}\ ,
\end{equation}
where we denoted irreducible representations of $\qgt$ by their 
dimensions (this is unambigous for this case). So the general 
solution can be written as
\begin{equation}
\R =\sum_{{\bf R}={\bf 1},{\bf 7},{\bf 14},{\bf 27}}A_{\bf_R}(x,q)
{\cal P}_{\bf R}\ ,
\end{equation}
where ${\cal P}_{\bf R}$ denotes the projector onto the subspace 
corresponding to $\bf R$ and $A_{\bf_R}(x,q)$ are scalar functions.
The construction of the projectors is described briefly in Appendix A.

Note that since $h_0=-2h_1-3h_2$ (corresponding to the fact that the 
level of the representation is $0$), the intertwining equation for $h_0$ 
is satisfied automatically. Therefore we are left 
with the task of solving the intertwining equations corresponding to the 
step operators for the affine root $\alpha_0$:
\begin{eqnarray}
\R(q^{-h_0/2}\otimes e_0+xe_0\otimes q^{h_0/2})=
(xq^{-h_0/2}\otimes e_0+e_0\otimes q^{h_0/2})\R\ ,\nn 
\R(q^{-h_0/2}\otimes f_0+\frac{1}{x}f_0\otimes q^{h_0/2})=
(\frac{1}{x}q^{-h_0/2}\otimes f_0+f_0\otimes q^{h_0/2})\R\ .\nn
\end{eqnarray}
Once one of these equations is solved, the other one is satisfied 
automatically due to symmetry reasons. The solution of these equations 
(as well as the projectors) were computed using the computer algebra 
program MAPLE and is given by
\begin{eqnarray}
&&A_{\bf 1}(x,q)=\rampl{2/3}\rampl{4}\nn
&&A_{\bf 7}(x,q)=\rampl{8/3}\nn
&&A_{\bf 14}(x,q)=\rampl{2/3}\nn
&&A_{\bf 27}(x,q)=1
\label{amplitudes}\end{eqnarray}
(where an overall normalization was chosen, since from the intertwining 
equation only the ratios of the amplitudes can be computed).

In (\ref{amplitudes}) and also in several later expressions we encounter 
powers of $q^{1/3}$. One has to choose an appropriate branch of the 
third root function to ensure the analytic dependence of the amplitudes 
on the coupling constant. Let us make the choice
\begin{equation}
q^{1/3}=\exp\left(\frac{4\pi^2 i}{3\beta^2}\right)\ ,
\end{equation}
and all powers of the form $q^{n/3}$ must be understood as the $n$th 
power of $q^{1/3}$, whenever $n$ is an integer. This choice of the 
branch is justified later by the consistency of the results we 
obtain for the $S$-matrix. One can see that at the point 
$\beta^2=8\pi /3$ the scattering becomes trivial.

To verify the 'brute force' calculation used to obtain the projectors 
and the amplitudes, we can use the tensor product graph method
\cite{tenspg}. 
In the above case, it says that the ratio of the amplitudes has to be 
\begin{eqnarray}
\frac{A_{{\bf R}_2}(x,q)}{A_{{\bf R}_1}(x,q)}=\left\langle \frac 12 
\left( C({{\bf R}_1})- C({{\bf R}_2}) \right)\right\rangle\ ,\nn
\langle{a}\rangle=\ampl{a}\ ,
\label{tpg}\end{eqnarray}
whenever the representation ${\bf R}_2$ occurs in the tensor product of 
${\bf R}_1$ with the representation corresponding to the affine root, 
which in our case is the adjoint representation $\bf 14$. $C({\bf R})$ 
denotes the value of the quadratic Casimir of the classical (i.e.
undeformed) 
group in the representation $\bf R$. Using the decomposition rules of
$\qgt$ 
and the values of the Casimir
\begin{equation}
C({\bf 1})=0\ , \ 
C({\bf 7})=4\ , \ 
C({\bf 14})=8\ , \ 
C({\bf 27})=28/3\ ,
\end{equation} 
it is easy to see that \rf{amplitudes} and \rf{tpg} are consistent. We 
remark that the quantum $R$-matrix of $\qgta$ has already appeared in  
connection with an exactly solvable $173$-vertex model on the lattice 
\cite{kuniba}.

In (\ref{amplitudes}) the following pole singularities occur:
\begin{itemize}
\item{} $x=q^4$: the singular piece is proportional to the singlet
projector. 
This corresponds to the occurence of the breathers as bound states of 
fundamental kinks.
\item{} $x=q^{4/3}$: the pole is in the $\bf 7$ channel, which gives 
higher kink multiplets and also the fundamental kink occuring as a 
bound state of fundamental kinks.
\item{} $x=q^{2/3}$: this gives another bound state in the
singlet+adjoint
representation.
\end{itemize}
\subsection{$S$-matrix for the fundamental solitons}
The $S$-matrix can be built from $\R$ using the principles of unitarity
and 
crossing symmetry. $\R$ satisfies the following relations:
\begin{eqnarray}
&&\R{\widehat{\cal R}}(1/x,q)=\id\ ,\label{unitarity}\\
&&\R(q^4/x,q)\frac{(x-1)(x-q^{10/3})(x-q^{4/3})q^{4/3}}
{(x-q^4)(x-q^{2/3})(x-q^{8/3})} \nn
&&=(C\otimes \id)(P_{12}\R)^{t_1}(C\otimes\id)P_{12}\label{crossing}\
,\nn
&&C=\left(\matrix{
0 & 0 & 0 & 0 & 0 & 0 & -q^{5/3} \cr
0 & 0 & 0 & 0 & 0 & q^{4/3} & 0 \cr
0 & 0 & 0 & 0 & -q^{1/3} & 0 & 0 \cr
0 & 0 & 0 & 1 & 0 & 0 & 0 \cr
0 & 0 & -q^{-1/3} & 0 & 0 & 0 & 0 \cr
0 & q^{-4/3} & 0 & 0 & 0 & 0 & 0 \cr
-q^{-5/3} & 0 & 0 & 0 & 0 & 0 & 0 \cr
}\right)\ .\ \label{chconj}
\end{eqnarray}
\rf{unitarity} is the unitarity property, while \rf{crossing} shows that 
$\R$ is almost crossing symmetric. The fact that the crossing
transformation
is given by $\theta \rightarrow i\pi -\theta$, is consistent with the 
following rapidity dependence of $x$
\begin{equation}
x=\exp\left[\left( \frac{4\pi}{\beta^2}h-h^\vee\right)\theta\right]=
\exp\left(\frac{2\pi}{\xi}\theta\right)\ ,
\end{equation}
where $h=4$ abd $h^\vee=6$ are the Coxeter and dual Coxeter numbers of 
$g_2$. The matrix $C$ given by \rf{chconj} is the charge conjugation
matrix 
in the homogeneous gradation. We remark that one has to be 
careful with the above notion of unitarity 
because it is not the usual notion of quantum field theory (QFT). 
Indeed, the $d_4^{(3)}$ affine Toda field theory is not expected 
to be a unitary field theory. This nonunitarity is usually 
reflected in the incorrect sign of the residues at the bound state
poles, 
but there may be even more severe violations of QFT unitarity, connected 
with a notion of ``pseudo-unitarity'' in a theory with indefinite 
metric on the space of states \cite{takacs}. 
In this paper, the term ``unitarity'' will 
mostly refer to relations of the type (\ref{unitarity}).

The universal R-matrix has the following additional properties:
\begin{eqnarray}
&&\left({\cal P}_{\bf 1}\right)_{12}
{\cal R}_{13}(xq^2,q){\cal R}_{23}(x/q^2,q)
\left({\cal P}_{\bf 1}\right)_{12}=\nn
&&\frac{(xq^{2/3}-1)(x-q^{4/3})(xq^2-1)}{(x-q^{2/3})(xq^{4/3}-1)(x-q^2)}
\left({\cal P}_{\bf 1}\right)_{12}\otimes {\bf I}_3
\ ,\label{brfuse}\\
&&\left({\cal P}_{\bf 7}\right)_{12}
{\cal R}_{13}(xq^{4/3},q){\cal R}_{23}(x/q^{4/3},q)
\left({\cal P}_{\bf 7}\right)_{12}=\nn
&&\frac{(xq^{4/3}-1)(x-q^{2/3})}{(x-q^{4/3})(xq^{2/3}-1)}
{\cal R}_{(12)3}(x,q)\ ,
\label{hkfuse}\end{eqnarray}
where $(12)$ denotes the seven-dimensional irreducible subspace in the 
tensor product of the first and second spaces.
(\ref{brfuse}) gives the bootstrap relation for the kink--breather and 
(\ref{hkfuse}) for the kink--higher kink scattering amplitudes. These 
formulae were computed by directly performing the matrix multiplications 
in MAPLE. They can be checked in the following easy way: from the left 
hand side of these equations, one can get the crossing transformation 
properties and the allowed poles of the right hand side expressions, 
which agree with the ones derived directly from the right hand side. 

We can change the gradation by applying the following similarity 
transformation
\begin{equation}
A \rightarrow x^{\frac{3}{4}h_1+\frac{5}{4}h_2}A
x^{-\frac{3}{4}h_1-\frac{5}{4}h_2},\ A=h_i,\ e_i,\ f_i,
\end{equation}
and, correspondingly
\begin{equation}
P_{12}\R_{spin}=x_1^{\frac{3}{4}h_1+\frac{5}{4}h_2}\otimes
x_2^{\frac{3}{4}h_1+\frac{5}{4}h_2}P_{12}\R
x_1^{-\frac{3}{4}h_1-\frac{5}{4}h_2}\otimes 
x_2^{-\frac{3}{4}h_1-\frac{5}{4}h_2}
\end{equation}
In this way we end up with the spin gradation, in which we have a charge 
conjugation matrix of the form
\begin{eqnarray}
C_{spin}=\left(\matrix{
0 & 0 & 0 & 0 & 0 & 0 & -1\cr
0 & 0 & 0 & 0 & 0 & 1 & 0 \cr
0 & 0 & 0 & 0 & -1& 0 & 0 \cr
0 & 0 & 0 & 1 & 0 & 0 & 0 \cr
0 & 0 & -1& 0 & 0 & 0 & 0 \cr
0 & 1 & 0 & 0 & 0 & 0 & 0 \cr
-1& 0 & 0 & 0 & 0 & 0 & 0 \cr
}\right)\ .
\end{eqnarray}
It can be checked readily that the evaluation representation in the 
homogenous gradation is given by 
\begin{eqnarray}
\pi^{(\theta)}_{spin}(e_0)=x^{1/4}e_0\ ,\ 
\pi^{(\theta)}_{spin}(e_1)=x^{1/4}e_1\ ,\ 
\pi^{(\theta)}_{spin}(e_2)=x^{1/12}e_2\ ,\nn
\pi^{(\theta)}_{spin}(f_0)=x^{-1/4}f_0\ ,\ 
\pi^{(\theta)}_{spin}(f_1)=x^{-1/4}f_1\ ,\ 
\pi^{(\theta)}_{spin}(f_2)=x^{-1/12}e_2\ ,\nn 
\end{eqnarray}
and therefore the spins are
\begin{eqnarray}
s_0=s_1=\frac{4\pi}{\beta^2}-\frac{3}{2}\ ,\ 
s_2=\frac{4\pi}{3\beta^2}-\frac{1}{2}\ .
\end{eqnarray}
So, in this gradation, the spins of the currents are the 'physical' ones 
as derived from the Lagrangian \rf{atftlagr} using only the canonical 
commutation relations and the canonical energy-momentum tensor.
Therefore 
for the theory \rf{atftlagr} the spin gradation can be considered to be 
the relevant one. Note that in this gradation the charge conjugation 
matrix is coupling constant independent and does not 
contain any phases (only some signs).

To get the real $S$-matrix, we must modify $\R_{spin}$ to be really
crossing 
symmetric, while preserving unitarity. Let us remember that the solution 
to the intertwining equations was determined only up to a scalar
function 
multiple. We can see that 
\begin{equation}
S(\theta)=\R_{spin}S_0(x)
\end{equation}
will be crossing symmetric and unitary if $f(x)$ satisfies
\begin{eqnarray}
&&S_0(x)S_0(1/x)=1\ , \nn
&&S_0(q^4/x)=S_0(x)\frac{(x-1)(x-q^{10/3})(x-q^{4/3})q^{4/3}}
{(x-q^4)(x-q^{2/3})(x-q^{8/3})}\ ,
\label{crosseq}\end{eqnarray}
which can be rewritten as
\begin{eqnarray}
&&S_0(\theta)S_0(-\theta)=1\ ,\nn
&&S_0(i\pi -\theta)=\nn
&&\frac
{\sinh\frac{\pi}{\xi}\left(\theta-i\pi\right)
\sinh\frac{\pi}{\xi}\left(\theta-\frac{2i\pi}{3}\right)
\sinh\frac{\pi}{\xi}\left(\theta-\frac{i\pi}{6}-\frac{i\xi}{2}
\right)}
{\sinh\frac{\pi}{\xi}\theta
\sinh\frac{\pi}{\xi}\left(\theta-\frac{i\pi}{3}\right)
\sinh\frac{\pi}{\xi}\left(\theta-\frac{5i\pi}{6}-\frac{i\xi}{2}
\right)}
S_0(\theta)\ .
\label{crosseq1}\end{eqnarray}

The solution to the equations (\ref{crosseq1}) is not unique. We will
choose 
a 'minimal' solution which means that  we allow only poles corresponding
to 
some degeneration of the 
$S$-matrix to a projector and their crossing symmetric partners (which 
correspond to the complementary projector). The unique function with 
this property is
\begin{eqnarray}
&&S_0(\theta )=
\prod _{k=0}^{\infty }
\frac
{\ggblock{1} \ggblock{\frac{2\pi}{\xi}} 
\ggblock{\frac{\pi}{3\xi}}\ggblock{1+\frac{5\pi}{3\xi}}
\ggblock{\frac{1}{2}+\frac{7\pi}{6\xi}}\ggblock{\frac{1}{2}+\frac{5\pi}{6\xi}}
}
{\ggblock{1+\frac{\pi}{\xi}} \ggblock{\frac{\pi}{\xi}} 
\ggblock{\frac{4\pi}{3\xi}} \ggblock{1+\frac{2\pi}{3\xi}}
\ggblock{\frac{1}{2}+\frac{\pi}{6\xi}}\ggblock{\frac{1}{2}+\frac{11\pi}{6\xi}}
}\ ,
\label{S0prod}\end{eqnarray} 
where
\begin{equation}
\left(x\right)=
\frac{\Gamma\left(x+\frac{2k\pi}{\xi}+\frac{i\theta}{\xi}\right)}
{\Gamma\left(x+\frac{2k\pi}{\xi}-\frac{i\theta}{\xi}\right)}
\end{equation}
 
$S_0(\theta )$ can be given an integral representation as follows:
\begin{equation}
S_0(\theta )=\exp\left[-i\int\limits_{-\infty}^{+\infty}
\frac{\sin k\theta \sinh \frac{1}{3}\pi k
\left(\cosh\left(\frac{\pi}{6}-\frac{\xi}{2}\right)k
-\frac{1}{2}\right)}
{k\sinh{\frac{\xi k}{2}}\cosh\frac{\pi k}{2}}dk
\right]
\label{S0int}\end{equation}
The transformation between the infinite product (\ref{S0prod}) and 
the integral (\ref{S0int}) can be performed using the formula
\begin{equation}
\ln \Gamma(z)=\int\limits_0^\infty\left[ (z-1)e^{-t}+
\frac{e^{-tz}-e^{-t}}{1-e^{-t}}\right]\frac{dt}{t}\ ,
\end{equation}
which is valid for $\Re e\ z >0$. This completes 
the derivation of the kink-kink scattering matrix.

\subsection{Breather-kink and breather-breather scattering amplitudes}

From the infinite product expansion (\ref{S0prod}) one can identify 
the poles and zeros of the scalar factor $S_0(\theta )$ and one can 
compute the pole structure of the total $S$-matrix with the help 
of (\ref{amplitudes}).
The bound state poles of the kink-kink $S$-matrix turn out to be:

\begin{itemize}
\item{} $\theta=i(\pi-m\xi)$, $m> 0$ integer: singlet breathers. The 
$m=1$ case will be denoted by $B_1$; the $m>1$ cases will be considered 
to be excited states of this breather and denoted by $B_1^{(m-1)}$.
\item{} $\theta=i\left(\frac{2\pi}{3}-m\xi\right)$, $m\geq 0$ integer: 
higher kink multiplets and the fundamental kink occuring as a 
bound state of fundamental kinks. We label them by $K_{m+1}$.
\item{}
$\theta=i\left(\frac{\pi}{6}-\left(m+\frac{1}{2}\right)\xi\right)$, 
$m\geq 0$ integer: bound states in the singlet+adjoint representation, 
denoted by $A_{m+1}$.
\end{itemize}
Of course, there are also the images of the poles required by crossing 
symmetry.

Using the formulae (\ref{brfuse}), (\ref{hkfuse}) and the integral 
formula for the scalar factor $S_0$ (\ref{S0int}), 
the S-matrices for the first 
two types of bound states can be computed. (The computation is much more 
difficult for the third type, although in principle straightforward).
Here we list only the amplitudes for the lowest lying breather with 
the fundamental kink and with itself:
\begin{eqnarray}
&&S_{K_1B_1}=\left\langle \frac{\pi}{2}+\frac{\xi}{2}\right\rangle_{K_1}
\left\langle \frac{5\pi}{6}-\frac{\xi}{2}\right\rangle_{K_2}
\left\langle -\frac{2\pi}{3}\right\rangle_{CDD_1}\nn
&&S_{B_1B_1}=\left\langle \xi\right\rangle_{B_1^{(1)}}
\left\langle \frac{2\pi}{3}\right\rangle_{B_1}
\left\langle -\frac{\pi}{3}+\xi \right\rangle_{B_2}\times\nn
&&\left\langle \frac{\pi}{6}-\frac{\xi}{2}\right\rangle_{AB_1}
\left\langle -\frac{\xi}{2}-\frac{\pi}{6}\right\rangle_{CDD_2} ,
\label{brS}\end{eqnarray}
where we used the notation
\begin{equation}
\langle x \rangle =\frac{\sinh\theta+i\sin x}{\sinh\theta-i\sin x}\ .
\end{equation}
The labels at the bottom of the blocks denote the bound states to which 
the corresponding poles belong. $K_1$ denotes the 
fundamental kink, and $B_1$ is the first breather originating from the 
pole in $S_{K_1K_1}$ at $\theta=i\pi-i\xi$. 
$B_1^{(1)}$ is the excited state of this first breather, corresponding 
to the pole in $S_{K_1K_1}$ at $\theta=i\pi-2i\xi$. $B_2$ is a singlet 
bound state of two $B_1$'s (a higher breather).
The pole ${AB_1}$ is presumably a singlet bound state pole (a breather) 
originating from the first adjoint+singlet 
soliton multiplet, which we denoted by $A_1$. This is supported by the 
observation that for $\xi > \frac{\pi}{3}$ when $A_1$ is not in the 
physical region, the pole $AB_1$ goes out of the physical strip as well.

What about the block labelled with $CDD_1$? To understand the 
notation, suppose that $\xi > \frac{\pi}{3}$. In this case the 
$A_1$ pole is missing in the $S_{K_1K_1}$ amplitude. 
Since all higher kink scattering matrices are proportional to the 
fundamental one, we get that there are only bound states in the 
singlet and fundamental representation. The spectrum very closely 
resembles that of the Zhiber-Mikhailov-Shabat (ZMS) model
\cite{smirnov}, 
with the difference that $d_4^{(3)}$ 
Toda theory has a seven-component kink instead of a three-component one. 

The block $CDD_1$ is just the factor which makes 
all the difference between the kink-breather 
scattering amplitude of the ZMS model and the kink-breather 
scattering amplitude of the 
$d_4^{(3)}$ Toda theory. This block only has poles outside the 
physical strip and so it is just an `innocent' CDD factor. 
The block indexed with  $CDD_2$ in $S_{B_1B_1}$ is similar in that the
poles 
originating from this factor are outside 
the physical strip for all values of $\xi$ for which the first 
breather pole exists, i.e. for $\xi <\pi$. Note that if 
$\xi >\pi$, the spectrum of the theory simplifies to the fundamental
kink
$K_1$.

The masses of the above particles are the following:
\begin{eqnarray}
&&m_{K_1}=m\ ,\
m_{K_{n+1}}=2m\sin\left(\frac{\pi}{6}+n\frac{\xi}{2}\right)\ , \ 
n\geq 1\ ,\nn
&&m_{B_1}=2m\sin\frac{\xi}{2}\ ,\ m_{B_1^{(n)}}=2m\sin\frac{(n+1)\xi}{2}
\ ,\ n\geq 1\ ,\nn
&&m_{B_2}=4m\sin\frac{\xi}{2}\sin\left(\frac{\pi}{3}+\frac{\xi}{2}\right)\
,\nn
&&m_{A_n}=2m\cos\left(\frac{\pi}{12}-(2n-1)\frac{\xi}{4}\right)\ ,\nn 
&&m_{AB_1}=4m\cos\left(\frac{\pi}{12}-\frac{\xi}{4}\right)\sin\frac{\xi}{2}\
.
\end{eqnarray}
The mass formula for $m_{AB_1}$ is analogous to those of the breathers 
orginating from the kinks $K_1$ and $K_2$.

To close the bootstrap for the generic case would require the 
amplitude for $A_1$. However, this multiplet is reducible, so 
there may be various mixings between the singlet and adjoint 
component, making it impossible to apply the simple-minded 
tensor product graph approach which is used to compute $R$-matrices 
in multiplicity-free representations. Therefore we do not 
investigate the bootstrap any further. We remark, however, that 
the spectrum is in agreement with the counting of the classical 
soliton solutions \cite{classol}, which shows that there should be 
two different soliton species associated to the two fundamental 
representations of $g_2$, i.e. to the representation $\bf 7$ and 
to the adjoint representation $\bf 14$. 
The above results imply that the adjoint soliton 
multiplet is extended by a singlet at the quantum level.
Furthermore, the mass ratio of the solitons $K_1$ and $A_1$ agrees 
with the first-order corrections computed in \cite{unstable}. 
The higher multiplets $K_n$ and $A_n$ (for $n>1$) can be thought 
of as excited states of the fundamental solitons in analogy with 
the ideas presented in \cite{excited}.

\subsection{Connection to real coupling affine Toda field theory}

Continuing the coupling $\beta$ to imaginary values we get an 
affine Toda field theory with real coupling. By the particle -- first 
breather correspondence principle, the $S$-matrix of the first breather 
must then become the amplitude for the fundamental particle of the 
real coupling case. 

The real coupling $d_4^{(3)}$ and $g_2^{(1)}$ affine Toda field 
theories form a dual pair. The $S$-matrix was computed  
on the basis of the hypothesis of `floating masses' and strong--weak 
coupling duality \cite{floating,duality1,duality2}. 
The $S$-matrix of the fundamental particle in $d_4^{(3)}$ can be 
written as
\begin{equation}
S_{11}=\left\langle \frac{2\pi}{H}\right\rangle
\left\langle \frac{2\pi}{3}\right\rangle
\left\langle \frac{2\pi}{H}-\frac{\pi}{3} \right\rangle
\left\langle \frac{\pi}{3}-\frac{4\pi}{H}\right\rangle
\left\langle -\frac{4\pi}{H}\right\rangle ,
\label{realS}\end{equation}
where
\begin{equation}
H=6\frac{8\pi+3\beta'^2}{4\pi+3\beta'^2}\ ,
\end{equation}
and $\beta'$ is the (real) coupling constant.

Putting $\beta'=i\beta$ we find the relation
\begin{equation}
\xi =\frac{\pi}{3}-\frac{4\pi}{H},
\end{equation}
and one can easily check that $S_{11}$ in (\ref{realS}) and 
$S_{B_1B_1}$ in (\ref{brS}) become identical. This is a
strong evidence that the $S$-matrix computed above really 
corresponds to the imaginary coupled $d_4^{(3)}$ affine 
Toda field theory. The second particle of the real coupling theory 
can be identified with $AB_1$, which was conjectured to be the lowest 
breather of the second soliton species. This is in accord with the 
assignment of the classical particles to the nodes of the Dynkin diagram 
of $g_2$.
 
\section{Perturbed $WA_2$ minimal models}

The action (\ref{atftlagr}) can be rewritten as the action of a
conformal 
$A_2$ Toda theory with a perturbation term:

\begin{eqnarray}
&&S=S_{A_2}+S_{pert}\ ,\nn
&&S_{A_2}=\int d^2x
\frac{1}{2}\partial_\mu\vec{\Phi}\partial_\mu\vec{\Phi}
+\frac{\lambda}{2\pi}\int d^2x \sum\limits_{j=0}^{1}
\exp\left(i\beta\frac{2}{(\vec{\alpha}_j,\vec{\alpha}_j)}
\vec{\alpha}_j\cdot\vec{\Phi}\right)\ ,\nn
&&S_{pert}=\frac{\lambda}{2\pi}\int d^2x 
\exp\left(i\beta\frac{2}{(\vec{\alpha}_2,\vec{\alpha}_2)}
\vec{\alpha}_2\cdot\vec{\Phi}\right)\ .
\end{eqnarray}

$S_{A_2}$ corresponds to a $WA_2$-invariant conformal field theory with 
the central charge \cite{bilgerv}
\begin{equation}
c=2\left(1-12\left(\frac{\beta}{\sqrt{4\pi}}
-\frac{\sqrt{4\pi}}{\beta}\right)^2\right)\ .
\end{equation}
When 
\begin{equation}
\frac{\beta}{\sqrt{4\pi}}=\sqrt{\frac{p}{p'}}\ ,\ p,\ p'\ 
{\rm coprime}\ {\rm integers}
\end{equation}
this is just the central charge corresponding to the $(p,p')$ minimal
model 
of the $WA_2$ algebra, which we denote with $WA_2(p,p')$. The field
content 
of the minimal model can be described by giving the spectrum of the
primary 
fields. These are labelled by four integers $n_1,\ n_2,\ m_1,\ m_2$ and 
denoted $\Phi(n_1n_2|m_1m_2)$. To each of these fields one can associate 
a vector 
\begin{equation}
\vec{\beta}(n_1n_2|m_1m_2)=\sum\limits_{i=1}^2 \left( \alpha_- (1-n_i)
+\alpha_+ (1-m_i\right)\omega_i\ , 
\end{equation}
where $\omega_i$ are the fundamental weights of $A_2$ and we define
\begin{equation}
\alpha_+^2=\sqrt{\frac{p}{p'}}\ ,\ \alpha_-=-\frac{1}{\alpha_+}\ ,\ 
\alpha=\alpha_+ +\alpha_-\ .
\end{equation}
The conformal weight of the field $\Phi(n_1n_2|m_1m_2)$ is \cite{fatluk}
\begin{equation}
h(n_1n_2|m_1m_2)=\frac{1}{2}\vec{\beta}^2-\alpha\vec{\rho}\vec{\beta}\ ,
\end{equation}
where
\begin{equation}
\vec{\rho}=\sum\limits_{i=1}^2 \omega_i\ .
\end{equation}
In our case the perturbing term turns out to be the field $\Phi(11|14)$.
The weight of this field is 
\begin{equation}
h(11|14)=6\frac{p}{p'}-3\ .
\end{equation}
To get a massive field theory, we require the perturbing field to be 
a relevant one, which means that its weight is less than one. Then 
we obtain the following condition:
\begin{equation}
\frac{p}{p'}<\frac{2}{3}\ {\rm or}\ \beta^2<8\pi /3
\label{relevcond}\end{equation}
This is only statisfied for nonunitary minimal models of $WA_2$ 
($|p-p'|>1$). 

For the above choice of $\beta$ the parameter $q$ becomes a root of
unity.
In the sine-Gordon and the ZMS model for these values of the couplings 
the state of space can be RSOS restricted to obtain integrable 
perturbations of the corresponding Virasoro minimal model
($\Phi_{(1,3)}$ 
perturbations from sine-Gordon 
\cite{smirresh}, and $\Phi_{(1,2)}$ \cite{smirnov} or $\Phi_{(1,5)}$ 
\cite{takacs} 
perturbations from the ZMS case). Following this analogy we assume 
that there exists a consistent restriction of the affine 
Toda field theory to the perturbed minimal model, which is described 
by the corresponding restriction of the representation theory of 
${\cal U}_q(A_2)$ at this value of $q$. In view of the above 
discussion, we cannot expect to get a unitary quantum field theory after 
the restriction. 

The condition (\ref{relevcond}) is in accord with what was said 
in connection with the special point $\beta^2=8\pi /3$ in subsection
2.1. If $\beta^2>8\pi /3$, the perturbation is irrelevant and flows 
back to the $WA_2$ minimal model in the infrared. In this regime 
the unrestricted theory is expected to be equivalent to two free 
bosons. 
Using the analogy with the Virasoro case we think that the restriction 
to the perturbed minimal model is equivalent to the Feigin-Fuchs 
free field construction of $WA_2(p,p')$ if the coupling of 
the perturbing term is set to zero.

To have this restriction we need to go over to a gradation in which 
all the rapidity dependence is carried by the generators corresponding 
to the root $\alpha_2$. In this gradation there is a ${\cal U}_q(A_2)$ 
algebra acting on the space of states. The fundamental kink-kink 
scattering matrix can then be decomposed into irreducible parts under 
${\cal U}_q(A_2)$. This is described in details in Appendix B.

One can see that for $q^6=1$ the singlet-singlet 
amplitude $S_{{\bf 1}{\bf 1}}^{{\bf 1}{\bf 1}}$ 
is unitary in itself. For generic $q$, 
the unitarity equations demands the presence of the particles 
in $\bf 3$ and $\bf\bar 3$ as intermediate states. Therefore we 
can interpret this fact as the existence of a consistent restriction to 
the singlet sector. The corresponding perturbed minimal models are 
$WA_2(3,p')+\Phi(11|14)$. Using the expression for 
$\R_{{\bf 1}{\bf 1}}^{{\bf 1}{\bf 1}}$ (\ref{decompose}), 
we conjecture that the scattering 
in these models is obtained from the following fundamental amplitude:
\begin{eqnarray}
\frac{\sinh\frac{\pi}{\xi}\left(\theta+i\pi\right)}
{\sinh\frac{\pi}{\xi}\left(\theta-i\pi\right)}
\frac{\sinh\frac{\pi}{\xi}\left(\theta+i\frac{2\pi}{3}\right)}
{\sinh\frac{\pi}{\xi}\left(\theta-i\frac{2\pi}{3}\right)}S_0(\theta )\ ,
\end{eqnarray}
where 
\begin{equation}
\xi=\frac{3\pi}{2p'-9}\ .
\end{equation}
The models $WA_2(3,p')+\Phi(11|14)$ are the analogs of the 
perturbed Virasoro minimal models $Vir(2,p')+\Phi(1,5)$. In the 
latter case, the state space is again restricted to allow only 
singlet representation and from the kink transforming in the 
singlet+doublet representation only the singlet piece survives
\cite{takacs}.

\section{Conclusions}

We have obtained the exact $S$-matrix of the imaginary coupled 
$d_4^{(3)}$ affine Toda field theory using the quantum affine symmetry 
of the model. This $S$-matrix proved to be consistent with 
the real coupling case using the breather-particle correspondence 
principle. It was argued that restricting the $S$-matrix in the 
$A_2$ gradation one should obtain exact $S$-matrices of 
the perturbed conformal field theories $WA_2(p,p')+\Phi(11|14)$. 
In the $p=3$ case the fundamental amplitude was computed explicitely. 
We saw that the relevance of the perturbation restricts $p,\ p'$ in 
such a way that unitary minimal theories are excluded.

To go further and obtain the $S$-matrix for the generic 
$WA_2(p,p')+\Phi(11|14)$ case, one would need an explicit expression 
for the  $6-j$--symbols of the algebra ${\cal U}_q(A_2)$, which, 
to our knowledge, is not yet available in general. 

On the other hand, one can get unitary perturbations of $WA_2(p,p')$ 
by considering the operator $\Phi(11|12)$. This operator has conformal 
weight
\begin{equation}
h(11|12)=\frac{4}{3}\frac{p}{p'}-1\ ,
\end{equation}
and so it is relevant for the unitary case $|p-p'|=1$. The corresponding 
affine Toda field theory is $g_2^{(1)}$ which has a symmetry algebra 
${\cal U}_q(d_4^{(3)})$. The kinks in this case are in the 
representation $\bf 8$ of the ${\cal U}_q(a_2)$ subalgebra of 
${\cal U}_q(d_4^{(3)})$. Unfortunately, while this representation 
is irreducible, taking the tensor product of $\bf 8$ with itself 
$\bf 8$ occurs twice, making the representation theoretic computation 
of the $S$-matrix extremely difficult.

A motivation for computing these $S$-matrices is the fact that 
the restrictions to perturbed $WA_2$ minimal models can be 
checked using the Thermodynamical Bethe Ansatz (TBA) and Truncated 
Conformal Space Approach (TCSA). Although there are still some 
obstacles in the way to a real check, this is an interesting line 
of research to pursue, since it could provide a verification of 
$S$-matrices for imaginary-coupled nonsimply-laced affine Toda 
field theories. 

On the semiclassical level, in such theories there are not 
enough solitons to fill up the affine algebra multiplets completely
\cite{mcghee}. 
It was proposed that the semiclassical spectrum must be extended 
in the quantum field theory to form complete multiplets 
\cite{An1smat,berlecl}. 
A combined TCSA/TBA verification of the $S$-matrix proposed in this 
paper could lend strong support to this idea.

\vspace{.5in}
\begin{center}
{\bf Acknowledgements}
\end{center}
I would like to thank G.M.T. Watts (King's College, London) for the 
discussions which directed my attention to the problem and also for 
the useful conversations (most of the time via e-mail) we had during 
the course of this work. I am also grateful to Z. Horv\'ath and L. Palla 
(Institute for Theoretical Physics, Budapest) for their help given 
at various stages of this work.
\vspace{.3in}
\begin{center}
{\Large\bf Appendix}
\end{center}
\vspace{.1in}
\appendix
\section{The invariant projectors}
The calculation of the quantum group invariant projectors ${\cal P}_{\bf
R}$ 
proceeds as follows. First we take the $49$-dimensional space of the 
tensor product ${\bf 7}\otimes{\bf 7}$ of two fundamental
representations.
We can define the action of ${\cal U}_q(G_2)$ on this space using the 
coproduct:
\begin{equation}
H_i=\Delta(h_i),\ E_i=\Delta(e_i),\ F_i=\Delta(f_i),\ i=1,2.
\end{equation}
Then we can find the highest weight vectors by computing the
intersection 
of the kernels of $E_i,\ i=1,2$. This gives a four-dimensional space, 
which can be spanned by the highest weight vectors of the
representations
${\bf 1},\ {\bf 7}, {\bf 14}$ and ${\bf 27}$. These will be labelled 
by $\repvect{1}{0}{0}$, $\repvect{7}{0}{1}$, $\repvect{14}{1}{0}$ and 
$\repvect{27}{2}{0}$, respectively. The first number is the dimension 
of the representation, while the second two numbers are the Dynkin 
indices of the corresponding weight vector of the Lie algebra $G_2$.
With these notations, we obtain the following expressions for the basis 
vectors of the representations  ${\bf 1},\ {\bf 7}$ and ${\bf 14}$:
\begin{eqnarray}
&&\repvect{14}{1}{0}\ ,\nn
&&\repvect{14}{-1}{3}=F_1\repvect{14}{1}{0}\ ,\nn
&&\repvect{14}{0}{1}=F_2\repvect{14}{-1}{3}\ ,\nn
&&\repvect{14}{1}{-1}=F_2\repvect{14}{0}{1}\ ,\nn
&&\repvect{14}{-1}{2}=F_1\repvect{14}{1}{-1}\ ,\nn
&&\repvect{14}{2}{-3}=F_2\repvect{14}{1}{-1}\ ,\nn
&&\repvect{14}{0}{0;1}=F_2\repvect{14}{-1}{2}\ ,\nn
&&\repvect{14}{0}{0;2}=F_1\repvect{14}{2}{-3}\ ,\nn
&&\repvect{14}{-2}{3}=F_1\repvect{14}{0}{0;1}\ ,\nn
&&\repvect{14}{1}{-2}=F_2\repvect{14}{0}{0;1}\ ,\nn
&&\repvect{14}{-1}{1}=F_2\repvect{14}{-2}{3}\ ,\nn
&&\repvect{14}{0}{-1}=F_2\repvect{14}{-1}{1}\ ,\nn
&&\repvect{14}{1}{-3}=F_2\repvect{14}{0}{-1}\ ,\nn
&&\repvect{14}{-1}{0}=F_1\repvect{14}{1}{-3}\ ,\nn
&&\repvect{7}{0}{1}\ ,\nn
&&\repvect{7}{1}{-1}=F_2\repvect{7}{0}{1}\ ,\nn
&&\repvect{7}{-1}{2}=F_1\repvect{7}{1}{-1}\ ,\nn
&&\repvect{7}{0}{0}=F_2\repvect{7}{-1}{2}\ ,\nn
&&\repvect{7}{1}{-2}=F_2\repvect{7}{0}{0}\ ,\nn
&&\repvect{7}{-1}{1}=F_1\repvect{7}{1}{-2}\ ,\nn
&&\repvect{7}{0}{-1}=F_2\repvect{7}{-1}{1}\ ,\nn
&&\repvect{1}{0}{0}\ .
\end{eqnarray}
(There are two vectors $\repvect{14}{0}{0}$, distinguished by a further 
degeneracy index.)

The invariant inner product on the representation space is defined by
the 
conditions:
\begin{equation}
E_{1,2}=F_{1,2}^\dagger\ ,H_{1,2}=H_{1,2}^\dagger\ ,
\end{equation}
and can be calculated as usual on complex vector spaces, but the 
conjugate of $q$ must be understood by treating $q$ as a formal
variable, 
i.e. $q^*=q$. With respect to this product, we can compute 
${\cal P}_{\bf R}$ for ${\bf R}={\bf 1},\ {\bf 7},\ {\bf 14}$ as 
the orthogonal projector on the corresponding subspace, while 
${\cal P}_{\bf 27}$ can be computed as the complementary projector (the 
irreducible subspaces are mutually orthogonal). 

All these projectors commute with the quantum group action and therefore 
they provide us with the four linearly independent solutions of the 
intertwining equations for ${\cal U}_q(G_2)$.

\section{The ${\cal U}_q(A_2)$ decomposition of the $S$-matrix}

The transformation to the $A_2$ gradation is described by the formula
\begin{equation}
P_{12}\R_{A_2}=x_1^{h_1+2h_2}\otimes
x_2^{h_1+2h_2}P_{12}\R
x_1^{-h_1-2h_2}\otimes 
x_2^{-h_1-2h_2}\ .
\end{equation}
In the $A_2$ gradation, the evaluation representation takes the form
\begin{eqnarray}
\pi^{(\theta)}_{A_2}(e_0)=e_0\ ,\ 
\pi^{(\theta)}_{A_2}(e_1)=e_1\ ,\ 
\pi^{(\theta)}_{A_2}(e_2)=x^{1/3}e_2\ ,\nn
\pi^{(\theta)}_{A_2}(f_0)=f_0\ ,\ 
\pi^{(\theta)}_{A_2}(f_1)=f_1\ ,\ 
\pi^{(\theta)}_{A_2}(f_2)=x^{-1/3}e_2\ ,\nn 
\end{eqnarray}
so now we have a genuine ${\cal U}_q(A_2)$ algebra acting on the 
representation space $\bf 7$, which decomposes into 
${\bf 1}\oplus{\bf 3}\oplus{\bf\bar 3}$ under this action.

Using this result, one can decompose $\R_{A_2}$ into irreducible 
components. The notation $\R_{{\bf a}{\bf b}}^{{\bf c}{\bf d}}$
means the component which maps from ${\bf c}\otimes{\bf d}$ to 
${\bf a}\otimes{\bf b}$. The result is the following:
\begin{eqnarray}
&&\R_{{\bf 1}{\bf 1}}^{{\bf 1}{\bf 1}}=\nn
&&{\frac{\left(x-q^{2}\right)\left(x^{2}q^{2/3}+x-2\,xq^{2/3}+xq^{2}-
2\,q^{8/3}x+q^{10/3}x-2\,q^{14/3}x+q^{16/3}x+q^{14/3}\right )}
{\left(x-q^{2/3}\right)\left(x-q^{4}\right)\left(x-q^{8/3}\right)}}\nn
&&\R_{{\bf 1}{\bf 3}}^{{\bf 1}{\bf 3}}=
\R_{{\bf\bar 3}{\bf 1}}^{{\bf \bar 3}{\bf 1}}=
{\frac {x^{2/3}\left(1-q^{4/3}\right)
\left(x-q^{2/3}+q^{4/3}-q^{2}\right)}
{\left(x-q^{2/3}\right)\left(x-q^{8/3}\right)}}\nn
&&\R_{{\bf 1}{\bf\bar 3}}^{{\bf 1}{\bf\bar 3}}=
\R_{{\bf 3}{\bf 1}}^{{\bf 3}{\bf 1}}=
{\frac {x^{1/3}\left(1-q^{4/3}\right)
\left(x-xq^{2/3}+xq^{4/3}-q^{2}\right)}
{\left(x-q^{2/3}\right)\left(x-q^{8/3}\right)}}\nn
&&\R_{{\bf 3}{\bf 1}}^{{\bf 1}{\bf 3}}=
\R_{{\bf 1}{\bf 3}}^{{\bf 3}{\bf 1}}=
\R_{{\bf\bar 3}{\bf 1}}^{{\bf 1}{\bf\bar 3}}=
\R_{{\bf\bar 1}{\bf 3}}^{{\bf\bar 3}{\bf 1}}=
{\frac{\left(x-q^{2}\right)\left(x-1\right)q^{2/3}}
{\left(x-q^{2/3}\right)\left(x-q^{8/3}\right)}}\nn
&&\R_{{\bf 3}{\bf\bar 3}}^{{\bf 1}{\bf 1}}=
\R_{{\bf 1}{\bf 1}}^{{\bf\bar 3}{\bf 3}}
{\frac{x^{1/3}\left(x-1\right)
\left(q^{4/3}-1\right)
\left(x-xq^{2/3}+xq^{4/3}-q^{10/3}\right )q^{4/3}}
{\left(x-q^{2/3}\right )\left (x-q^{4}\right )\left(x-q^{8/3}\right
)}}\nn
&&\R_{{\bf\bar 3}{\bf 3}}^{{\bf 1}{\bf 1}}=
\R_{{\bf 1}{\bf 1}}^{{\bf 3}{\bf\bar 3}}=
{\frac{ x^{2/3}\left(x-1\right)
\left (q^{4/3-1}\right)
\left (x-q^{2}+q^{8/3}-q^{10/3}\right)q^{4/3}}
{\left(x-q^{2/3}\right)\left(x-q^{4}\right)\left(x-q^{8/3}\right)}}\nn
&&\R_{{\bf 3}{\bf 3}}^{{\bf 3}{\bf 3}}=
\R_{{\bf\bar 3}{\bf\bar 3}}^{{\bf\bar 3}{\bf\bar 3}}=
{\cal P}_{\bf 6}-
{\frac{q^{8/3}x-xq^{2}+xq^{2/3}-x-xq^{4/3}+q^{4/3}+x^{2}q^{2}-q^{10/3}x}
{\left(x-q^{2/3}\right)\left(x-q^{8/3}\right)}}{\cal P}_{\bf 3}\nn
&&\R_{{\bf 3}{\bf\bar 3}}^{{\bf 3}{\bf\bar 3}}=
{\frac{x^{2/3}\left(1-q^{2/3}\right)}{x-q^{2/3}}}{\cal P}_{\bf 8}+
{\frac{x^{2/3}\left(q^{4/3}-1\right)
f_1(x,q)}
{\left(x-q^{4}\right)\left(x-q^{2/3}\right)\left(x-q^{8/3}\right)}}
{\cal P}_{\bf 1}\nn
&&\R_{{\bf\bar 3}{\bf 3}}^{{\bf\bar 3}{\bf 3}}=
{\frac {{x^{1/3}}\left(1-q^{2/3}\right)}
{x-q^{2/3}}}{\cal P}_{\bf 8}+
{\frac {{x^{1/3}}\left(1-q^{4/3}\right)
f_2(x,q)}
{\left(x-q^{4}\right)\left(x-q^{2/3}\right)\left(x-q^{8/3}\right)}}
{\cal P}_{\bf 1}\nn
&&\R_{{\bf 3}{\bf\bar 3}}^{{\bf\bar 3}{\bf 3}}=
\R_{{\bf\bar 3}{\bf 3}}^{{\bf 3}{\bf\bar 3}}=
\frac{q^{1/3}(x-1)}{x-q^{2/3}}{\cal P}_{\bf 8}+
{\frac{\left(x-1\right)f_3(x,q)}
{\left(x-q^{4}\right)\left(x-q^{2/3}\right)\left(x-q^{8/3}\right)}}
{\cal P}_{\bf 1}
\label{decompose}\end{eqnarray}
where we used the notation
\begin{eqnarray}
&&f_1(x,q)=-q^{6}+q^{16/3}-q^{14/3}x+xq^{4}
-q^{10/3}x+q^{10/3}x^{2}+q^{8/3}-\nn
&&q^{8/3}x^{2}-q^{2}+x^{2}q^{2}+xq^{2}-2\,xq^{4/3}+q^{4/3}+xq^{2/3}-x\nn
&&f_2(x,q)=xq^{6}-q^{16/3}x-q^{14/3}x^{2}+
2\,q^{14/3}x+q^{4}x^{2}-xq^{4}-q^{4}+\nn
&&q^{10/3}-q^{10/3}x^{2}-q^{8/3}+q^{8/3}x-xq^{2}+xq^{4/3}-x^{2}q^{2/3}+x^{2}\nn
&&f_3(x,q)=( q^{8/3}x^{2}+q^{14/3}x-xq^{6}+q^{8/3}x
+xq^{4/3}-2\,xq^{4}-2\,xq^{2}-\nn
&&x+q^{10/3}x+q^{10/3} )q^{2/3}\ .
\end{eqnarray}

${\cal P}_{\bf 1}$ and ${\cal P}_{\bf 8}$ denote the projectors on the 
irreducible subspaces $\bf 1$ and $\bf 8$, while ${\cal P}_{\bf 3}$ 
projects on $\bf 3$ or $\bf\bar 3$ and ${\cal P}_{\bf 6}$ 
on $\bf 6$ or $\bf\bar 6$ depending on which occurs in the decomposition 
of the subspace indicated in the indices of $\R$.

\end{document}